\documentclass[fleqn,usenatbib]{mnras}

\usepackage{newtxtext,newtxmath}

\usepackage[T1]{fontenc}
\usepackage{ae,aecompl}

\usepackage{graphicx}	

\usepackage{amsmath}	
\usepackage{amssymb}	
\usepackage{comment}
\usepackage{siunitx}
\usepackage{booktabs}

\usepackage{xcolor}

\def\be{\begin{equation}}
\def\ee{\end{equation}}
\newcommand\code[1]{\textsc{\MakeLowercase{#1}}}

\newcommand\quotes[1]{``{#1}"}
\def\gsim{\lower.5ex\hbox{\gtsima}} 
\def\lsim{\lower.5ex\hbox{\ltsima}} 
\def\gtsima{$\; \buildrel > \over \sim \;$} 
\def\ltsima{$\; \buildrel < \over \sim \;$} \def\gsim{\lower.5ex\hbox{\gtsima}} 
\def\lsim{\lower.5ex\hbox{\ltsima}} 
\def\simgt{\lower.5ex\hbox{\gtsima}} 
\def\simlt{\lower.5ex\hbox{\ltsima}}

\def\msun{{\rm M}_{\odot}}

\def\cc{\rm cm^{-3}}
\def\kms{\,\rm km\,s^{-1}}

\def\fesc{f_{\rm esc}}

\def\S*{$\Sigma_{\rm SFR}$}

\def\CII{\hbox{[C~$\scriptstyle\rm II $]~}}

\def\OIII{\hbox{[O~$\scriptstyle\rm III $]~}}

\def\HI{\hbox{H~$\scriptstyle\rm I\ $}}

\def\CIIion{\hbox{C~$\scriptstyle\rm II $~}}

\definecolor{apcolor}{HTML}{b3003b}
\definecolor{afcolor}{HTML}{800080}
\definecolor{lvcolor}{HTML}{DF7401}
\definecolor{mdcolor}{HTML}{01abdf} 
\definecolor{cbcolor}{HTML}{ff0000}
\definecolor{sccolor}{HTML}{cc5500} 
\definecolor{sgcolor}{HTML}{00cc7a}

\renewcommand{\d}{\mathrm{d}}

\let\oldnabla\nabla
\renewcommand{\nabla}{\vec{\oldnabla}}

\graphicspath{{plots/}} 

\defcitealias{chevalier_clegg:1985}{CC85}
\def\citeCC85{\citetalias{chevalier_clegg:1985}}
\defcitealias{Fujimoto:2020qzo}{F20}
\def\citeF20{\citetalias{Fujimoto:2020qzo}}
\defcitealias{Pizzati20}{P20}
\def\citeP20{\citetalias{Pizzati20}}
\defcitealias{chevalier_clegg:1985}{CC85}
\def\citeCC85{\citetalias{chevalier_clegg:1985}}

\title[~\CII halos in ALPINE galaxies]{[CII] halos in ALPINE galaxies: smoking-gun of galactic outflows?}

\author[Pizzati et al.]{E. Pizzati$^{1,2}$\thanks{\href{mailto:elia.pizzati@sns.it}{elia.pizzati@sns.it}},
A. Ferrara$^{1}$,
A. Pallottini$^{1}$,
L. Sommovigo$^{1}$,
M. Kohandel$^{1}$,
S. Carniani$^{1}$
\\
$^{1}$ Scuola Normale Superiore, Piazza dei Cavalieri 7, 56126 Pisa, Italy\\
$^{2}$ Leiden Observatory, Leiden University, P.O. Box 9513, 2300 RA Leiden,
The Netherlands\\
}

\date{Accepted XXX. Received YYY; in original form ZZZ}

\pubyear{2022}

\begin{document}
\label{firstpage}
\pagerange{\pageref{firstpage}--\pageref{lastpage}}
\maketitle

\begin{abstract}
ALMA observations have revealed that many high redshift galaxies are surrounded by extended ($10-15$ kpc) \CII-emitting halos which are not predicted by even the most advanced zoom-in simulations. Using a semi-analytical model, in a previous work we suggested that such halos are produced by starburst-driven, catastrophically cooling outflows. Here, we further improve the model and compare its predictions with data from 7 star-forming ($10\lsim \rm SFR/\msun\,yr^{-1}\lsim 100$) galaxies at $z=4-6$, observed in the ALPINE survey. We find that (a) detected \CII halos are a natural by-product of starburst-driven outflows; (b) the outflow mass loading factors are in the range $4 \simlt \eta \simlt 7$, with higher $\eta$ values for lower-mass, lower-SFR systems, and scale with stellar mass as $\eta \propto M_*^{-0.43}$, consistently with the momentum-driven hypothesis. Our model suggests that outflows are widespread phenomena in high-$z$ galaxies. However, in low-mass systems the halo extended \CII emission is likely too faint to be detected with the current levels of sensitivity.   
\end{abstract}




\section{Introduction} \label{sec:introduction}

Investigating the complex environments of galaxies at the end of the Epoch of Reionization \citep[EoR, redshift $z\gtrsim 5.5$; for recent reviews see][]{Dayal:2018hft, hodge_dacunha} is one of the most pressing research goals of modern astrophysics. As shown by cosmological simulations \citep{fire2, Vogelsberger:2019ynw, sphinx, Pallottini22}, galaxies at the EoR show different properties with respect to the ones seen in the local Universe (e.g., smaller sizes, higher specific star formation rates, warmer dust). 

In the last few years, astounding progress has been made in detecting primordial galaxies up to $z\sim13$ \citep[][]{oesch_2011, harikane_2021}. This has fueled a strong and widespread interest in detailed studies of their internal structure. As a result, these early systems are now being routinely identified in large-scale optical/Near-Infrared (NIR) surveys  \citep{refId0, bouwens_15}, and targeted interferometric observations at Far-Infrared (FIR) wavelengths \citep{lefevre:2019, bouwens_rebels}.

Specifically, the Hubble Space Telescope (HST) measured sizes and morphological properties of high-redshift galaxies at rest-frame ultraviolet (UV) wavelengths \citep{oesch2009structure,Shibuya:2015qfa,bradley_2014, bouwens_15,oesch_2016, bouwens2017z,livermore_2017, kawamata2018size}. These studies successfully characterized the evolution of the rest-frame galaxy UV luminosity functions (LFs), star formation, stellar buildup history, and size growth, giving a first statistical characterization of galactic systems up to $z\approx 10$. 

At the same time, the appearance of the Atacama Large Millimeter/submillimeter Array (ALMA) and the NOrthern Extended Millimeter Array (NOEMA) have opened a new window on the primordial Universe, exploring the obscured star formation and interstellar medium (ISM) line emission at rest-frame FIR wavelengths up to $z\approx13$ \citep[][]{harikane_2021}. Combining the information coming from the dust continuum emission, as well as from some relevant FIR emission lines such as \CII $158 \,\mu\mathrm{m}$, \OIII $88 \,\mu\mathrm{m}$, and CO from various rotational levels, several works studied the internal properties of galaxies, such as their assembly history, ISM thermal structure, gas dynamics, dust/metal enrichment, and interstellar radiation field \citep[ISRF,][]{maiolino2015,capak2015,pentericci2016, matthee2017, carniani2018,Hashimoto2018, gallerani:2018, carniani2020, calura2021}.

In particular, the \CII $^2P_{3/2} \rightarrow ^2P_{1/2}$ fine-structure transition has been the workhorse for many observational and theoretical endeavors. Due to its brightness and ubiquity \citep[it traces many different phases of the ISM, e.g.,][]{stacey1991,hollenbach1999,wolfire2003}, this line can provide essential information on the physics of galaxies at high redshift \citep[e.g.,][]{vallini2015, olsen2018}. For instance, different works have focused on interpreting the local \CII - SFR relation \citep[][]{delooze, hc15, hc18} in the context of early systems \citep[][]{carniani2018, schaerer2020, carniani2020}, studying its dependence on the burstiness of the star formation process, the gas density, and the intensity of the ISRF \citep[][]{ferrara:2019, pallottini:2019, vallini2020}. 
 
A key result that emerged thanks to spatially resolved observations of the \CII line is the existence of \textit{extended} \CII \textit{halos} around normal, star-forming galaxies at high redshift \citep[][]{Fujimoto19, ginolfi:2019, Fujimoto:2020qzo, herrera2021kiloparsec, fudamoto2022, akins22}. These halos imply the presence of singly ionized carbon extending out to distances significantly larger than the size of galactic disks. 

The first evidence for such halos was obtained by \citet{Fujimoto19} who stacked ALMA observations of 18 main-sequence galaxies at redshifts $z=5-7$ directly in the uv-visibility plane. The data have revealed the presence of a \CII surface brightness $\approx 5\times$ times more extended than the HST stellar continuum and ALMA dust continuum maps. This implies the presence of a carbon halo extending out to $ \approx 10\,\mathrm{kpc}$ from the stacked galaxy center (i.e., well into the circumgalactic medium of these galaxies). 

This early result was then backed up by detailed observations within the ALMA ALPINE Large Program \citep[][]{lefevre:2019, bethermin:2019, Faisst:2020}. This program opened up for the first time the possibility of studying extended halos around  normal, high-redshift galaxies at the individual level. \citet[][hereafter \citeF20]{Fujimoto:2020qzo} measured the physical extent of \CII line-emitting gas in 23 star-forming galaxies at $z=4-6$. This study found that, in the vast majority of cases, the size of the \CII emitting region exceeds the size of the UV stellar continuum by factors of $\sim 2-3$, and concluded that at least $\sim 30\%$ of the sources have a \CII halo extending over ten-kpc scales. Interestingly, the authors studied also the dependence of the size of this extended \CII line structure on few galactic properties, finding a positive correlation with e.g. star-formation rate (SFR) and stellar mass. Follow-up observations are needed to test whether \CII halos are a universal feature in high-redshift star-forming galaxies, and to investigate the observed spread in the halos size and morphology. 

On top of these findings, observations of SPD.81, a $z \approx 3$ gravitationally lensed dusty star-forming galaxy, showed that $\sim50\,\%$ of \CII emission arises in the external (i.e., not FIR-bright) region of the galaxy \citep{Rybak20}. Another interesting result came from deep observations of HZ4, a typical star-forming galaxy at $z\approx5.5$ \citep[][]{herrera2021kiloparsec}. These authors found evidence for a \CII emission extending beyond the dust and UV continuum disk, and forming a halo of $\approx6$ kpc in radius. 

More recently, further signatures of \CII halos have been found at high redshift ($z\sim7$), both in stacked and individual studies. Within the context of the REBELS survey \citep[][]{bouwens_rebels}, \citet[][]{fudamoto2022} observed a stacked \CII emission $\times\sim2$ larger than the dust continuum and the rest-frame UV ones, in agreement with the results of \citeF20 at $z\sim5$. This suggests that normal, star-forming galaxies present an extended \CII emission feature over a wide range of redshifts. Using individual observations of the galaxy A1689-zD1 ($z=7.13$), \citet{akins22} comes to the same conclusion by proving that the \CII line is extended up to a radius $r\sim12\,\mathrm{kpc}$, with the \OIII line and the UV-continuum not extending farther than $r\sim4\,\mathrm{kpc}$. 

The mounting observational evidence for the existence of these extended halos inevitably calls for a thorough theoretical investigation of their origin, properties and evolution. In principle, these issues could be clarified by zoom-in cosmological simulations. With their ability to resolve structures down to Molecular Clouds (MC) scales, they are well suited to study the internal structure of primordial galaxies and of their circumgalactic medium (CGM) \citep[e.g.,][]{pallottini:2019}. However -- as shown by \citet{Fujimoto19} -- mock observations generated by using $z=6$ zoom simulations \citep{pallottini2017b, Arata:2019} fail to reproduce the observed \CII surface brightness distribution of the emitting material in extended halos.
These independent studies agree in predicting a \CII emission that is slightly more extended than the stellar continuum but drops very rapidly at distances considerably smaller than $10\,\mathrm{kpc}$. The resulting profiles are characterized by a value of the surface brightness which, in the external regions, is at least one order of magnitude lower than observed. This mismatch between theory and observations could originate either from issues in modelling FIR emission lines or because of some key physical ingredients that are not well captured by the numerical implementations adopted in the cosmological simulations. In any case, such tension represents a serious challenge in our understanding of early galaxy formation. 

In order to bridge the gap between simulations and observational data, exploratory work can be useful to examine the physical implications of the observed emission, and guide the development of new simulations on the right track. In this context, \citet[][hereafter \citeP20]{Pizzati20} showed that the \CII halos observed by \citet{Fujimoto19} can be reproduced by assuming that they are created by past (or ongoing) galactic outflow activity.
In \citeP20, we developed semi-analytical model for a supernova-driven galactic wind, and computed the \CII emission expected to arise from the outflowing gas. We found very good agreement between the stacked \citet[][]{Fujimoto19} data and the model predictions, provided that: (a) the outflow mass loading factor $\eta = \dot M/{\rm SFR}$, where $\dot M$ is the outflow rate, is relatively high ($\eta \approx 3$); (b) the escape fraction of ionizing photons from the parent galaxy is low ($\fesc < 1\%$).
The first condition ensures that the density of the outflowing gas is high enough to match the observed \CII surface brightness; the second prevents the galaxy radiation from ionizing the gas in the halo and leaving only scarce amounts of \CIIion in the gas. For simplicity, we explored the two extreme cases $\fesc = 20\%$ (i.e., photons from the galaxy dominate the radiation budget) and $\fesc=0$ (i.e. the UV background is the dominant source of radiation in the halo), and find that only the latter scenario can explain the presence of the observed \CII extended emission.

Here, we elaborate further on the possible connection between outflow activity and \CII halos by significantly improving the \citeP20 model. In particular, we introduce a more realistic treatment of the gravitational influence exerted by the parent galaxy and its dark matter (DM) halo on the outflowing gas, and we account for the cosmic microwave background (CMB) suppression of the \CII line \citep[e.g.,][]{vallini2015}. We then compare the results of this updated model with the \CII surface brightness profiles observed by the ALPINE survey (\citeF20). We aim to study galaxies at the individual level, in order to characterize the properties of these system and statistically infer possible scaling relations present at high redshift. 

The paper is structured as follows. In Sec. \ref{sec:model}, we summarize the main features of the \citeP20 model, with particular emphasis on the improvements introduced in this work. Sec. \ref{sec:data} contains a brief overview of the ALPINE observational data used in this work. Sec. \ref{sec:data_analysis} and \ref{sec:results} focus on the comparison between model results and ALPINE data. These results are discussed and interpreted in the framework of current galaxy formation and evolution theories in Sec. \ref{sec:discussion}. Conclusions are given in Sec. \ref{sec_conclusions}.

\section{Outflow model} \label{sec:model}

The \citeP20 model makes quantitative predictions on the properties of a \CII halo generated by outflowing gas from a galaxy. It takes as inputs three key parameters: the outflow mass loading factor ($\eta$), the parent galaxy star formation rate ($\mathrm{SFR}$), and circular velocity of the hosting DM halo ($v_c$). Once these three parameters are set, the model can predict the spatial distribution of the \CII surface brightness produced by the outflowing gas. The profile can then be convolved with the same beam as the observations, and compared directly with data.

The model builds on previous studies of galactic winds: we consider the same setting as the pioneer work of \citet[][hereafter \citeCC85]{chevalier_clegg:1985}. The active star forming region of the galaxy is modeled as a sphere of radius $R=300\,\mathrm{pc}$. Energy and mass are uniformly injected from this region at a constant rate to mimic the input of supernovae (SN) associated with the star formation activity. This energy and mass deposition drives a spherically symmetric, hot, and steady wind that expands freely in the region outside the galaxy (the presence of the IGM/CGM is neglected).

As shown by e.g. \citet{Thompson16}, the assumption of adiabatic expansion in the original \citeCC85 model breaks down for high values of the outflow mass loading factor ($\eta \gtrsim 1$). For large $\eta$ values, then, the inclusion of radiative cooling results in a rapid decrease of the gas temperature, creating a cold/warm wind mode ($T\approx 10^{2-4}\,\mathrm{K}$) that propagates outward with lower velocity and higher density.
This physical process is so abrupt and dramatic that is known as \textit{catastrophic cooling}, and it has been advocated by both semi-analytical models and simulations as one of the main cold wind formation channels \citep{McCourt:2015, sarkar:2015, Thompson16, Scannapieco:2017, Schneider:2018, Gronke&Oh:2020, fielding2021}. In \citeP20, we focus on this cooling scenario, as the formation and survival of \CII in the gas is guaranteed only for relatively low temperature values ($T\lesssim 10^4\,\mathrm{K}$). 
Radiative losses are described in our wind equations by the net (i.e. cooling-heating) cooling function $\Lambda$, computed from the Tables in \citet{gnedin2012cooling}. We assume here solar metallicity for the gas, in accordance with the results of simulations at $z\approx6$ \citep{pallottini2017, Pallottini22} and with the extrapolation of the mass-metallicity relation at high redshifts \citep[][we are interested in the mass range $10^9-10^{11}\,\msun$]{mannucci:2012}. 

Under the hypothesis of photo-ionization equilibrium (PIE), the cooling (and heating) rates are strongly dependent on the ionizing radiation fields impinging on the gas. In turn, these fields can be produced either by leakage of ionizing photons from the parent galaxy, or by the cosmological contribution from all the other galaxies and quasars (the UV background, UVB). In \citeP20, we extensively discuss how the overall properties of the wind depend on the relative intensity of these two radiation fields. As already explained in the Introduction, models in which $\fesc$ is substantial, dramatically reduce the abundance of the CII ion by populating higher ionization states of the atom (CIII, CIV). Thus, we follow \citeP20 and fix $\fesc=0$, i.e. we consider only we consider the UVB as the only radiation field acting on the gas.

Along with cooling, it is important to consider the gravitational effects of the galaxy disk and DM halo on the gas, which are neglected in the original \citeCC85 formulation. This is an important effect: as the gas cools down radiatively, it slows down considerably, and thus its velocity becomes comparable with the escape velocity from the galactic system. In \citeP20, the effects of the DM halo gravitational potential on the gas are included by parametrizing them with the halo circular velocity $v_c$.

With these assumptions, a cool wind that drives carbon (and other metals) into the CGM of the galaxy is produced. The wind is slowed down by the gravitational potential, and reaches a stalling radius $r_\mathrm{stop}$ at distances comparable with the sizes of the observed \CII halo \citep[][]{Fujimoto19}. 
The hypothesis of photo-ionization equilibrium is then used to compute the abundance of neutral hydrogen and \CIIion in the gas\footnote{We use a carbon abundance $A_{\mathrm{C}} = 2.69 \times 10^{-4}$ \citep{asplund2009}.}. 
Together with the outflow temperature and density, relative abundances of $\HI$ and $\CIIion$ determine the emissivity of the \CII line. Considering that the line is typically optically thin, a straightforward line integration holds the \CII surface brightness distribution. Further details about the outflow model and \CII line emission predictions can be found in Sec. 4 and Sec. 5 of \citeP20, respectively. We now proceed on to detail two new specific improvements we have implemented with respect to \citeP20.

\subsection{Gravitational potential} \label{sec:gravity}

\begin{figure*}
	\centering
	\includegraphics[width=1.0\textwidth]{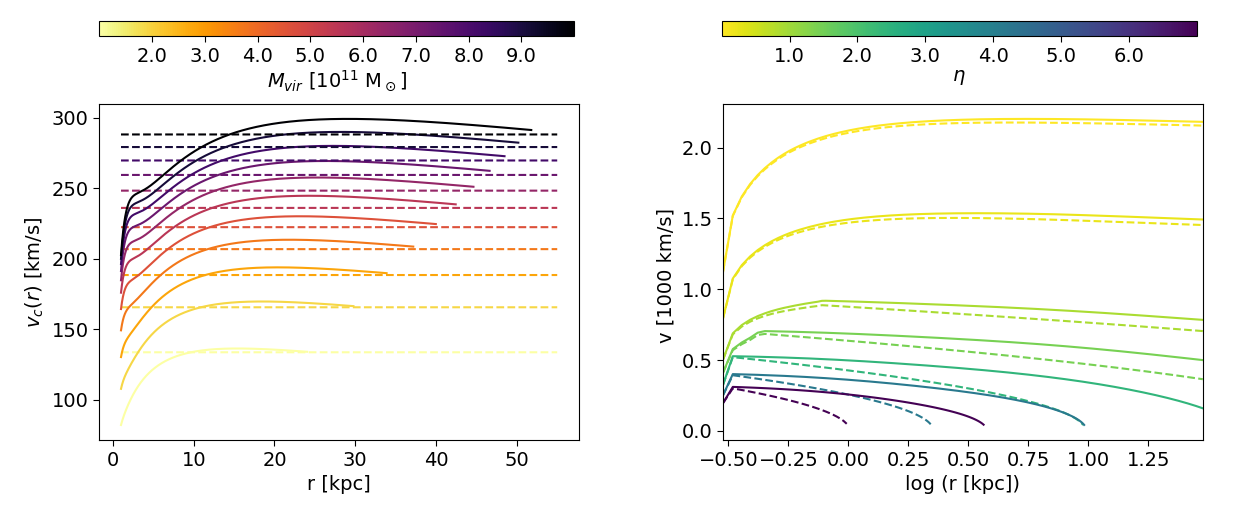}
	\caption{\textit{Left:}  Circular velocity, $v_c$, as a function of radius for a NFW halo density profile \citep[][]{NFW_profile} and a Sersic-like stellar distribution (solid lines). Different halos are color-coded according to their virial mass, $M_{\rm vir}$. The dashed lines represent the profiles for $v_c$ employed in \citeP20, and also the values of $V_c$ that are used as parameters in our model. \textit{Right}: Outflow velocity (note the different scale with respect to the left panel) for different values of the mass loading factor $\eta$. The other parameters are fixed to the fiducial values reported in the text. The NFW+bulge (isothermal sphere) case is shown with solid (dashed) lines.
	\label{fig:gravity}
	}
\end{figure*}

Gravity can significantly slow down the gas expansion. This effect can be accounted for by introducing a gravitational potential $\phi(r)$ produced by a matter density distribution $\rho(r)$. Under the assumption of spherical symmetry, we can write the radial gradient of the potential as:
\begin{align}
\frac{\d \phi(r)}{\d r}=\frac{G M(r)}{r^2} = \frac{v_c^2(r)}{r}.
\end{align}
In this expression, $M$ is the total mass contained in a sphere of radius $r$, and $v_c(r)$ is the local circular velocity:
\begin{align}
v_{c}(r) &= \sqrt{\frac{GM(r)}{r}}
\label{eq:v_c_definition_model}
\end{align}
 
The total mass is the sum of the DM halo contribution, $M_\mathrm{DM}(r)$, and of the central galaxy's baryonic component, $M_\mathrm{b}(r)$\footnote{We neglect outflow self-gravity}. In \citeP20, we included only the contribution from the DM halo (neglecting the presence of the galactic disk) by assuming a singular isothermal sphere, for which $\rho\propto r^{-2}$. Such choice is particularly convenient because it yields $M\propto r$ and $v_c(r) = {\rm const.}$ 
Assuming an isothermal profile was a reasonable approximation in \citeP20, as our aim was to compare our model only with stacked ALMA observations \citep[][]{Fujimoto19}. In this work, however, it becomes essential to model the effects of gravity in a realistic way to ensure a fair comparison between our model predictions and the individual galactic properties inferred from observations. 

For this reason, we assume here a Navarro-Frenk-White \citep[NFW,][]{NFW_profile} density profile for DM. Hence, the DM total mass $M_\mathrm{DM}(r)$ can be obtained analytically by integrating the density profile $\rho_\mathrm{NFW}(r)$ in a volume of radius $r$:
\begin{align}
M_\mathrm{DM}(r) = \frac{M_\mathrm{vir}}{\log(1+c) - \frac{c}{1+c}}\Bigg[\log\left(\frac{r_s + r}{r_s}\right) + \frac{r_s}{r_s + r} -1\Bigg] \end{align}
where $M_\mathrm{vir}$ ($r_\mathrm{vir}$) is the virial mass (radius) of the halo, $c$ is the concentration parameter, and $r_s = r_\mathrm{vir}/c$. Clearly, this is valid only for $r<r_\mathrm{vir}$, which is canonically defined as $r_\mathrm{vir} = (3M_\mathrm{vir}/4\pi \delta_c\Bar{\rho})^{1/3}$, where $\delta_c = 200$ is the mean halo overdensity with respect to the cosmic background density $\Bar{\rho}$ at redshift $z$. For the concentration parameter, $c(M_\mathrm{vir},z)$, we adopt the fit given by \citet{dutton2014cold}. 

We also include the baryonic contribution to the total mass. Consistently with our assumption of spherical symmetry for the system, we adopt a Sersic profile with $n=1$ (i.e., an exponential distribution) for the radial component of the stellar density:
\begin{align}
\rho_b (r) = \frac{M_*(M_\mathrm{vir}, z)}{8\pi r^3}\,e^{-r/r_*(r_\mathrm{vir})}, 
\end{align}
where $M_*$ is the total stellar mass and $r_*$ is the scale radius of the distribution. We link the halo mass $M_\mathrm{vir}$ and the stellar mass $M_*$ -- for a given redshift $z$ -- via the abundance matching approach of \citet[][]{behroozi2013average}. Further, we compute the radius $r_*$ a function of the virial radius by introducing the spin parameter $ \lambda = r_* / r_\mathrm{vir}$, which we set to $\lambda = 0.01$ \citep[][]{Shibuya:2015qfa}. By integrating the density $\rho_b(r)$, we can then find the baryonic mass inside a sphere of radius $r$, $M_b(r)$, and consequently the total mass distribution $M(r)$ and the circular velocity $v_c(r)$. Note that in this formulation we are not accounting for the contribution of gas to the baryonic mass. This component may be a significant fraction of the total baryonic mass \citep[e.g.,][]{dessauges2020}; however, it is largely unconstrained and to a first approximation we neglect its presence in the galaxy mass budget. 

The only parameters in this formulation are then the virial mass $M_\mathrm{vir}$ and the redshift $z$. Once they are set, we can characterize completely the halo gravitational potential $\phi(r)$ and study its effect on the outflow. In \citeP20, we used the (constant) circular velocity $v_c$ as a model parameter. Thus, here it is convenient to introduce a similar reparametrization of $M_\mathrm{vir}$ by defining the (dark-matter-only) circular velocity at the virial radius:
\begin{equation}
    V_c \equiv \sqrt{\frac{GM_\mathrm{vir}}{r_\mathrm{vir}}} \label{eq:v_circ}
\end{equation} 

In the left panel of Fig. \ref{fig:gravity}, we plot $v_c(r)$ as a function of the galactocentric radius $r$, and for different values of the virial mass $M_\mathrm{vir}$. Values of $v_c(r)$ are shown only for $r<r_\mathrm{vir}$, where our region of interest resides. The dark-matter-only, isothermal-sphere case adopted in \citeP20 is plotted for comparison with dashed horizontal lines. These lines also represent the value of the circular velocity at virial radius $V_c(M_\mathrm{vir}, z)$.
Fig. \ref{fig:gravity} (right panel) quantifies the impact of different density profiles on the outflow velocity, $v$. Again, solid lines refers to the model used in this work, while dashed lines show the results for the \citeP20 model. In these runs, the following fiducial values of the model parameters have been fixed: $\mathrm{SFR}=50\,\msun\,\mathrm{yr}^{-1}$, $V_c = 200\,\kms$, and $z=5$. The mass loading factor $\eta$ is varied in the range $\eta\approx 0.5-7$. As detailed in \citeP20, higher values of $\eta$ result in lower launching velocities for the wind, and ultimately in lower stalling radii. 

Adopting the NFW halo density profile considerably reduces the effects of gravity, especially for high values of $\eta$. Introducing the effects of baryonic matter increases the circular velocity in the central regions of the galactic system but it does not have a strong impact on the evolution of gas at large radii ($r\gtrsim 5\,\mathrm{kpc}$). As a result, the peak outflow velocity in the present is very similar to the one in \citeP20, but the subsequent decrease is less steep here. Thus, the stalling radius $r_\mathrm{stop}$ is larger for a fixed value of the mass loading factor $\eta$.

\subsection{CMB line suppression}\label{sec:CMB_suppression}

Several works have pointed out that, as redshift progressively increases, CMB photons start to play an important role in regulating the FIR lines emission \citep[][]{gong2012,dacunha2013,pallottini:2015, vallini2015, kohandel:2019}. In particular, at $z\approx4-6$, the temperature of CMB photons becomes high enough to produce an important effect on the overall \CII emission of the gas. This effect is twofold: on the one hand, the CMB radiation field affects the atomic level populations, and hence the \CII transition rate; on the other hand, the \CII emission is observed against the CMB uniform background, which must be subtracted out when computing the \CII surface density. 

In \citeP20, we ignored these effects for simplicity. However, as shown by e.g. \citet{kohandel:2019}, their role in suppressing \CII line emission in the external, low-density regions of high-$z$ may be non-negligible. Thus, we take these effects into account by introducing the ratio $\zeta$ between the specific intensity of the \CII emission observed against the CMB and the intrinsic one \citep[see e.g.,][]{dacunha2013}. This ratio can be written as \citep[][]{gong2012}:
\begin{equation}\label{eq:suppression}
    \zeta = 1-\frac{B_{\nu_*}(T_\mathrm{CMB}(z))}{B_{\nu_*}(T_\mathrm{exc})}\,,
\end{equation}
where $B_\nu$ is the black body intensity and $\nu_* \simeq 1901\,\mathrm{GHz}$ is the restframe frequency of the $\CII$ transition. The excitation temperature $T_\mathrm{exc}$ describes the effect of the CMB photons on the populations of the levels involved in the \CII transition (i.e., the \CIIion fine structure doublet, $^2P_{3/2}-\,^2P_{1/2}$). An expression for $T_\mathrm{exc}$ is obtained in Appendix \ref{sec:cmb_appendix} (eq. \ref{eq:exc_temp}).

\begin{figure*}
	\centering
	\includegraphics[width=1.0\textwidth]{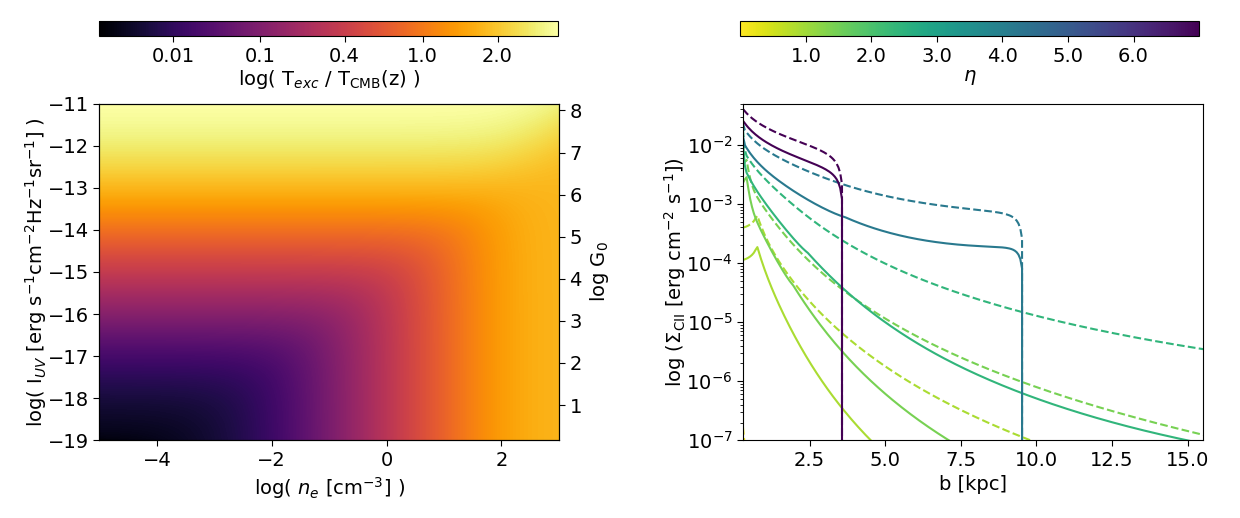}
	\caption{\textit{Left}: Heat map showing the ratio between the excitation temperature and the CMB one ($T_\mathrm{exc}/T_\mathrm{CMB}(z)$), computed at redshift $z=5$, as a function of the free electron density, $n_e$, and UV radiation field intensity, $I_{\rm UV}$ at $1330$ \AA. The latter quantity is also expressed in terms of the {Habing flux} $G_0$ (in units of the averaged Milky Way value $1.6\times 10^{-3}\,\mathrm{erg} \,\mathrm{cm}^{-2}\,\mathrm{s}^{-1}$; right axis). The gas temperature is assumed to be $T=10^3\,\mathrm{K}$, and the gas electron fraction is fixed to $x_e = n_e/n =0.5$. \textit{Right}: \CII surface density $\Sigma_\mathrm{CII}$ (eq. \ref{eq:final_intensity}) as a function of the impact parameter $b$, for different values of the mass loading factor $\eta$. Solid (dashed) lines show the results with (without) the inclusion of CMB suppression.
	\label{fig:cmb}
	}
\end{figure*}

In Fig. \ref{fig:cmb} (left panel), we show how the excitation temperature depends on two key parameters: the gas (electronic) density $n_e$, and the flux in the UV band -- in the range $-19 < \log_{10} \left(I_{\rm UV}[\mathrm{erg}\mathrm{s}^{-1}\mathrm{cm}^{-2}\mathrm{Hz}^{-1}\mathrm{sr}^{-1}]\right) < -11$. Specifically, we show the ratio between the excitation temperature and the CMB one, $T_\mathrm{exc}/T_\mathrm{CMB}(z=5)$. As such ratio approaches unity, the \CII emission becomes indistinguishable from the CMB background, and thus the final \CII flux is strongly suppressed. 

We can identify three different physical regimes determining $T_\mathrm{exc}$. For $n_e \gg 1 \, \cc$, collisions dominate and the excitation temperature approaches the gas kinetic temperature $T$; if a strong UV flux is illuminating the gas, the excitation temperature gets approximately equal to the UV color temperature ($T_\mathrm{UV}\approx 10^4\,\mathrm{K}$; see eq. \ref{eq:uv_color}); for low $n_e$ densities and weak UV fluxes, instead, CMB photons dominate the transition rates, and thus the excitation temperature approaches $T_\mathrm{CMB}$.

In order to account for CMB effects on \CII emission in our model, we insert the suppression factor $\zeta$ (eq. \ref{eq:suppression}) in the integral defining the \CII surface density. Exploiting the assumed spherical symmetry, we express $\Sigma_\mathrm{CII}$ as a function of a single variable, the impact parameter $b$ (i.e., the distance between the line of sight and the center of the galaxy), by integrating the \CII emissivity $\dot{\varepsilon}_\mathrm{CII} = n^2 \Lambda_{\mathrm{CII}}$ (eq. 17 in \citeP20) along the line of sight:
\begin{align}
\Sigma_\mathrm{CII}(b)&=\int_{-\infty}^{+\infty} \dot{\varepsilon}_\mathrm{CII}\zeta \d s = 2 \int_{b}^{+\infty} \dot{\varepsilon}_\mathrm{CII}(r)\zeta(r)\, \frac{r}{\sqrt{r^2 - b^2}}\d r
\label{eq:final_intensity}
\end{align}
This integral depends implicitly on the thermodynamic and ionization state of the outflow, as well as the ionizing radiation field contributions from the galaxy and the UVB.

As an illustration, in Fig. \ref{fig:cmb} we show the $b$-dependence of the surface brightness profiles as a function of the mass loading factor $\eta$. The other parameters are set to their fiducial values (see Sec. \ref{sec:gravity}).
From the Figure, we see that the suppression of the \CII emission can be significant. The effects are dramatic for low $\eta$ values ($\zeta \approx 0.05-0.3$), and more moderate (but still relevant) in the high $\eta$ range ($\zeta \approx0.3-0.8$). Therefore, we conclude that CMB suppression plays an important role in regulating the total \CII emission at high redshift, by diminishing the total \CII luminosity and favoring high gas densities in the outflow (or, equivalently, high values of the mass loading factor).
The surface density profiles shown in Fig. \ref{fig:cmb} are converted into fluxes per unit area (see eq. 24 in \citeP20) and convolved with the same beams as the observations by \citeF20. These final steps allow us to directly compare the model results to the observations. 

\section{ALPINE sample}\label{sec:data}

The ALMA ALPINE survey \citep[][]{lefevre:2019, Faisst:2020, bethermin:2019} targeted 118 normal, star-forming ($10\lsim \rm SFR/\msun\,yr^{-1}\lsim 100$) galaxies at $z\approx4-6$. It measured the spatial distribution of the \CII line ($64\%$ detection rate) and the (rest-frame) FIR continuum ($21\%$). On top of that, it exploited observations from HST, VLT, and Spitzer to build a multiwavelength catalog for these systems \citep[][]{Faisst:2020}. 

\citeF20 analyze ALPINE observations focusing on the extension of the \CII emitting galactic halos. They select the 23 galaxies from the ALPINE sample whose \CII emission is observed above the $5\sigma$ level, and they compare the (averaged) radial profile of the \CII line with the ones of the rest-frame UV (stellar) and FIR (dust) continuum. 

Out of these 23 galaxies, they identify 7 systems belonging to the \quotes{\CII halo} category. These are defined as systems whose peripheral region (i.e., the region obtained by masking the central galaxy contribution) shows a \CII line at $>4\sigma$ and the (rest-frame) UV and FIR continuum at $<3\sigma$. 
Note that such thresholds inevitably depend on the sensitivity of observations. New, deeper observations could in principle reveal the presence of \CII halos even in other ALPINE systems that were not inserted in the \quotes{\CII halo} category by \citeF20 \citep[for more details see][]{herrera2021kiloparsec}.
However, this classification is the first attempt to characterize \CII halos in individual high-$z$ galaxies and to study the statistical properties of their host systems. For this reason, the work of \citeF20 is ideally suited for comparing the outputs of our model with observational data. 

The 7 \quotes{\CII halo} galaxies identified in \citeF20 have redshifts in the range $z_\mathrm{CII}\approx4.5-5.5$, star formation rates SFR$ > 15\, \msun {\rm yr}^{-1}$, and stellar masses $M_* > 5 \times 10^9 \msun$. In Tab. \ref{tab:alpine_sample}, we report the value of $z_\mathrm{CII}$, $\mathrm{SFR}$, and $M_*$ inferred from observations. Redshifts are measured spectroscopically using the \CII line, while star formation rates and stellar masses are inferred from SED fitting of the multiwavelength ancillary data \citep[][]{Faisst:2020}. 

As detailed in Sec. \ref{sec:gravity}, we convert the values of the stellar masses inferred observationally into DM halo virial masses using the fitting functions obtained by \citet[][]{behroozi2013average}. We also compute the resulting circular velocity at virial radius from eq. \ref{eq:v_circ}. We obtain virial masses in the range $M_\mathrm{vir}\approx (3-7) \times 10^{11}\,\msun$ and circular velocities $V_{\rm c} \approx 200-250\,\kms$. For reference, these quantities are shown in Tab. \ref{tab:alpine_sample} together with the rest of observational data. 

\begin{table*}
\centering
\setlength{\extrarowheight}{3pt}
\begin{tabular}{ c | c c c c c  | c c c c }
\toprule
\multicolumn{1}{c}{}& \multicolumn{5}{c}{{Galaxy data}}
& \multicolumn{4}{c}{{Model Predictions}}\\
 ID  & $z_\mathrm{CII}$    & $\mathrm{SFR}$   & $M_*$   & $M_\mathrm{vir}$     & $V_{\rm c}$ &  $\mathrm{SFR}$  & $V_{\rm c}$  & $\eta$ & $\chi^2(\Theta_\mathrm{max})$/ndof \\ 

   &  &  [$\msun\,\mathrm{yr}^{-1}$]   & [$10^{9}\,\msun$]    &  [$10^{11}\,\msun$]     &  [$\mathrm{km}\,\mathrm{s}^{-1}$]  & [$\msun\,\mathrm{yr}^{-1}$]   & [$\mathrm{km}\,\mathrm{s}^{-1}$] &  &  \\ 
 \midrule
\midrule
DC$396844$    &    $4.54$    &    $55^{+40}_{-25}$    &   $7.3^{+2.6}_{-2.6}$   &   $4.4^{+2.8}_{-1.8}$    &   $210^{+90}_{-66}$  &
$64^{+51}_{-29}$    &   
$170^{+54}_{-46}$   &   
$5.1^{+2.2}_{-1.6}$   & $0.60$ \\
DC$630594$    &    $4.44$    &    $31^{+24}_{-15}$    &   $5.9^{+2.3}_{-1.7}$   &   $3.9^{+2.6}_{-1.5}$    &   $199^{+85}_{-59}$ &
$34^{+32}_{-16}$    &   
$196^{+43}_{-47}$   & 
$6.8^{+4.6}_{-2.4}$    &   $0.68$ \\
DC$683613$    &    $5.54$    &    $58^{+44}_{-26}$    &   $14.7^{+5.7}_{-4.4}$   &   $6.7^{+5.9}_{-3.7}$    &  $241^{+145}_{-102}$ & 
$84^{+60}_{-37}$    &  
$160^{+72}_{-44}$   &  
$4.8^{+2.0}_{-1.5}$     &    $0.45$ \\
DC$880016$    &    $4.54$    &    $32^{+25}_{-15}$    &   $5.6^{+2.3}_{-1.6}$  &   $3.7^{+2.6}_{-1.4}$    &   $197^{+85}_{-58}$  &
$37^{+30}_{-17}$    &  
$170^{+51}_{-40}$   &  
$5.5^{+2.5}_{-1.8}$    &    $0.97$ \\
DC$881725$    &    $4.58$    &    $88^{+61}_{-43}$    &   $9.1^{+4.0}_{-2.1}$   &   $4.9^{+3.5}_{-2.1}$    &   $217^{+98}_{-68}$  &
$102^{+50}_{-44}$    &   
$222^{+75}_{-82}$   &  
$4.3^{+3.2}_{-1.2}$  &   $0.68$ \\
VC$5100537582$    &    $4.55$    &   $15^{+14}_{-6}$    &   $5.7^{+2.0}_{-1.6}$& $3.8^{+2.6}_{-1.4}  $  &   $198^{+83}_{-58}$  & 
$23^{+15}_{-9}$    &  
$162^{+29}_{-27}$   &  
$6.9^{+2.7}_{-2.0}$   &   $1.58$ \\
VC$5110377875$    &    $4.55$    &    $99^{+79}_{-41}$    &  
 $14.7^{+4.7}_{-5.6}$   &   $6.8^{+5.1}_{-3.9}$    &    $243^{+125}_{-97}$&    
$107^{+78}_{-49}$    &   $241^{+63}_{-41}$   & 
$3.7^{+2.0}_{-1.1}$   &   $0.73$  \\
 \bottomrule
\end{tabular}
\caption{Properties of the \quotes{CII halo} sample taken from the \citeF20 study. From left to right: name of the ALPINE source (\quotes{DC} stands for \quotes{DEIMOS COSMOS}, while \quotes{VC} stands for \quotes{VUDS COSMOS}); values of redshift ($z_\mathrm{CII}$), star formation rate ($\mathrm{SFR}$), and stellar mass ($M_*$) taken from \citet[][]{Faisst:2020}; halo mass ($M_\mathrm{vir}$) and global circular velocity ($v_c$) inferred from \citet[][]{behroozi2013average}; results form our MCMC analysis, i.e., median and $1\sigma$ uncertainties of the posterior distributions for the three parameters considered in our model: $\eta$, $\mathrm{SFR}$, and $V_{\rm c}$; $\chi^2$ of the model-data comparison for the values of the parameters that maximize the posterior distribution, $\Theta_\mathrm{max}$, divided by the number of degrees of freedom (ndof$=6$). Uncertainties on the redshift measurements are very small and not shown here. 
\label{tab:alpine_sample}
}
\end{table*}

\section{Comparison with data} \label{sec:data_analysis}

We now proceed to compare our model predictions with the observed \CII profiles for the 7 systems presented in Sec. \ref{sec:data}. We take as parameters in our model the mass loading factor ($\log\eta$), the star formation rate ($\log\mathrm{SFR}$), and the circular velocity ($\log V_{\rm c}$). For every one of the systems considered, we assume flat priors on $\log \eta$, and use the information presented in Sec. \ref{sec:data} to set Gaussian priors on $\log\mathrm{SFR}$ and $\log V_{\rm c}$ (the median and variance are chosen according to the values reported on Tab. \ref{tab:alpine_sample}).
A detailed discussion of our methods is given in Appendix \ref{sec:data_analysis_details}. 

\begin{figure*}
	\centering
	\includegraphics[width=0.495\textwidth]{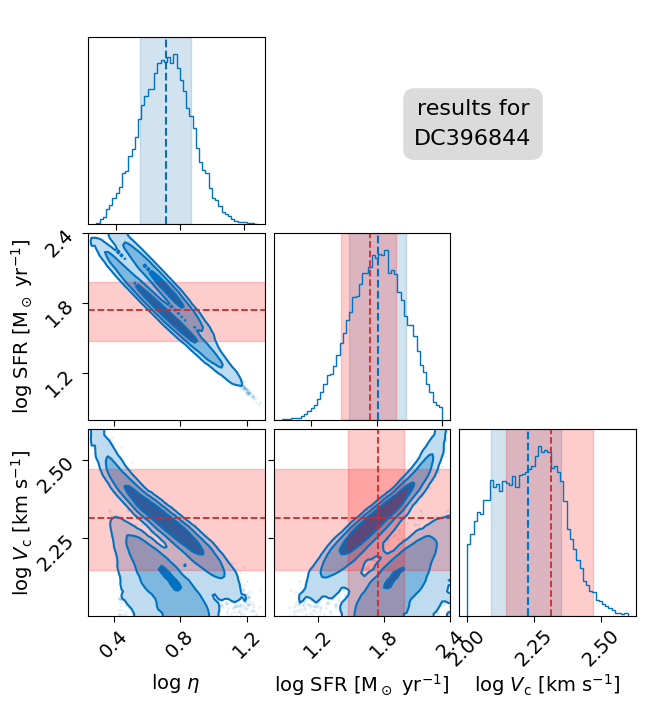}
	\includegraphics[width=0.495\textwidth]{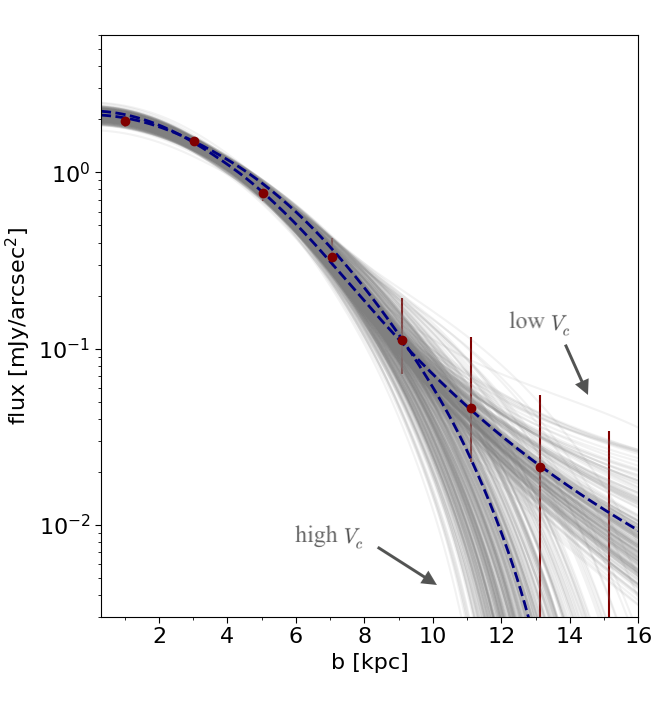}
	\caption{\textit{Left:}
	Corner plots of the 3D posterior distribution $P(\log\eta, \log\mathrm{SFR}, \log V_{\rm c}|d,m)$ for the galaxy DC$396844$.
	The values of the parameters $\mathrm{SFR}$ and $V_{\rm c}$ inferred from observations (Tab. \ref{tab:alpine_sample}) are shown with red dashed lines, together with their uncertainties (red shaded regions). The blue dashed vertical lines represent the median values of the parameters ($1\sigma$ errors are again shown as shaded regions).
	Similar plots for the full ALPINE \quotes{\CII halo} sample are shown in Fig. \ref{fig:corner_2}.
	\textit{Right}: Comparison of predicted \CII profiles (light gray lines) obtained for DC$396844$ with the observed one (\citeF20, red points). The theoretical profiles are generated by randomly sampling the posterior distribution $500$ times. The blue dashed lines correspond to the values of the parameters for which the posterior exhibits a local peak (see left panel).
	\label{fig:corner_1}
	}
\end{figure*}

We run a Markov-Chain Montecarlo (MCMC) using the Python package \code{emcee} to explore the posterior distributions. As an illustration, the corner plot in Fig. \ref{fig:corner_1} shows the 1D and 2D marginalized posterior for one of the galaxies in the sample, DC$396844$. The same plots for the other galaxies are shown in Appendix \ref{sec:other_sample}, Fig. \ref{fig:corner_2}. The predicted median values and $1\sigma$ errors of the different parameters are given on the right side of Tab. \ref{tab:alpine_sample}.

We also show the values of $\chi^2(\Theta_\mathrm{max})$ (computed at the peak of the posterior distribution, $\Theta_\mathrm{max}$) in the last column of Tab. \ref{tab:alpine_sample}. We divide these by the number of degrees of freedom for our 3-parameter model: $\mathrm{ndof}=6$. We find that the values of $\chi^2$/ndof are reasonably low, implying that our model reproduces observations quite well. This can be confirmed visually by the right panel of Fig. \ref{fig:corner_1}, where we plot the observational data (red points) along with synthesized emission profiles from our model (grey lines), created by randomly sampling the posterior distribution. 

\section{Results} \label{sec:results}

The corner plots shown in Fig.s \ref{fig:corner_1} and \ref{fig:corner_2} reveal that, for most sources, there is evidence for a bimodal shape of the posterior distribution. The origin of the bimodality can be understood focusing on the emission profiles for DC$396844$ (right panel of Fig. \ref{fig:corner_1}): the \CII profiles for low-$V_{\rm c}$ halos are even more extended than the observed ones; on the contrary, higher $V_{\rm c}$ values result in a more compact emission that is barely compatible with the observed ones. Given the large relative uncertainty on data points at large impact parameter, both behaviors are consistent with the data. 

In almost all cases, the recovered values for $\mathrm{SFR}$ and $V_{\rm c}$ (blue shaded regions in the corner plots) match very well the independent determinations from the SED fitting (red shaded regions). This is partly due to our choice for the prior distributions; even so, given the good agreement between our model and the data profiles, it signifies that well-matching solutions can be found for values of the star formation rates and stellar masses that are compatible with observational estimates.

The posterior distributions we obtain for the $\eta$ parameter are particularly interesting, as they represent one of the few (indirect) measurements of the mass loading factor available at the high redshift. As shown in Tab. \ref{tab:alpine_sample}, we infer relatively high median values for the mass loading factor, $4.3 \le \eta \le 6.8$. 

\begin{figure*}
	\centering
	\includegraphics[width=1.0\textwidth]{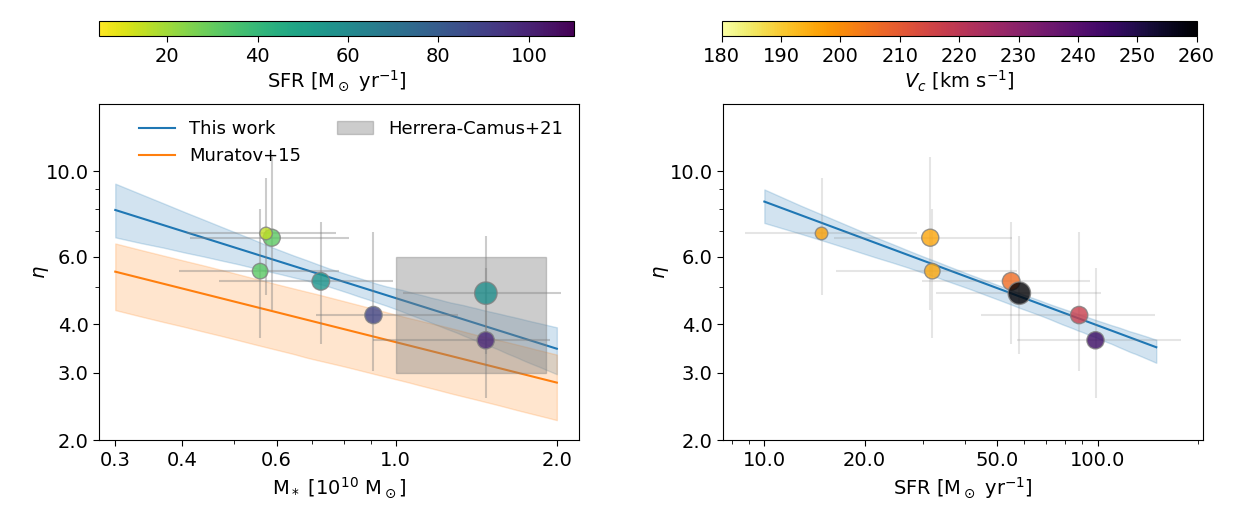}
    \caption{\textit{Left}: Dependence of the mass loading factor $\eta$ (shown are median values and associated 1$\sigma$ error) on the stellar mass $M_*$. Stellar masses are obtained from observational data (Table \ref{tab:alpine_sample}). The blue line shows the best power-law fit to the data. For reference we also show predictions from simulations \citep[][orange, eq. \ref{Muratov}]{muratov2015} and observations \citet[][grey rectangle]{herrera2021kiloparsec}. Data points are color-coded according to their SFR, as inferred from observations. The size of every point is proportional to the value of the likelihood $\mathcal{L}(\Theta_{\rm max})$ (eq. \ref{eq:likelihood}). \textit{Right}: Same as left panel, but for the star formation rate (SFR). The color of the data points corresponds to their virial velocity $V_{\rm c}$ (see colorbar), as described in Tab. \ref{tab:alpine_sample}.
    \label{fig:final}
 	}
\end{figure*}

As the mass loading factor depend on the star formation rate and stellar mass of galaxies, it is useful to plot the predicted values of $\eta$ as a function of these two physical properties. In Fig. \ref{fig:final} we plot the results and perform power-law fits, finding that both $\eta-M_*$ and $\eta-\mathrm{SFR}$ are inversely correlated. The best-fit power law scalings are:
\begin{align}
&\log \eta = (-0.43\pm0.14) \log(M_*/10^{10}\msun) +\log(4.7\pm0.3)\,\label{eq:eta_mstar}\\
&\log\eta = (-0.32\pm0.05) \log( \mathrm{SFR} /\msun\,\mathrm{yr}^{-1}) + \log(17.5\pm3.6) \label{eq:eta_sfr}
\end{align}
The dependence $\eta \propto M_*^{-0.43}$ is close to that expected from a momentum-driven outflow, for which $\eta \propto M_*^{-1/3}$ \citep[][]{Finlator:2007mh, dave2012}. This is consistent with the fact that outflowing gas in our model solutions loses much of its thermal energy via radiative cooling in a region very close to the launching site (Sec. \ref{sec:model}). The analogue dependence for the $\eta-\mathrm{SFR}$ is a consequence of the galaxy main sequence, where most of the ALPINE systems reside \citep[][]{Faisst:2020}.

Numerical simulations of starburst galaxies also find mass loading factors that are compatible with momentum-driven outflows. For instance, \citet{muratov2015} uses the \code{fire} zoom cosmological simulations \citep[][]{hopkins2014} to obtain the following power-law relation with stellar mass $M_*$:
\begin{equation}
\label{Muratov}
\log \eta = (-0.36\pm0.02) \log(M_*/10^{10}\msun) +\log(3.6\pm 0.7).
\end{equation}
We add this relation to our results in Fig. \ref{fig:final} for comparison. We find a perfectly compatible scaling, but our values of $\eta$ tend to be systematically higher; it is hard, however, to tell whether this difference ($\approx 1-2\sigma$) is due to some systematical bias (see Sec. \ref{sec:loading_factors} for more details) or it contains some interesting physical insight. 

Note that, in principle, the presence of additional energy sources such as Active Galactic Nuclei (AGN) may disrupt the simple power-law scaling between $\eta$ and $M_*$ \citep[see e.g.,][]{zhang2021empirical}. We do not expect, however, our sample to be affected by this, as galaxies in the ALPINE catalogue show no sign of AGN activity (Barchiesi et al. in prep., \citealt{2022arXiv220603510F}).

\section{Discussion}\label{sec:discussion}

In this work, we have applied the (improved) \citeP20 model to ALPINE systems showing signs of extended \CII emission, arguing that: (a) detected \CII halos are a natural by-product of starburst-driven outflows; (b) the mass loading factors of these outflows have values in the range $4 \simlt \eta \simlt 7$ and scaling compatible with the momentum-driven hypothesis. In the following, we further elaborate on this picture by investigating its consequences and comparing it with previous studies.

\subsection{Halo-outflow connection: observational evidence}

\begin{figure}
	\centering
	\includegraphics[width=0.48\textwidth]{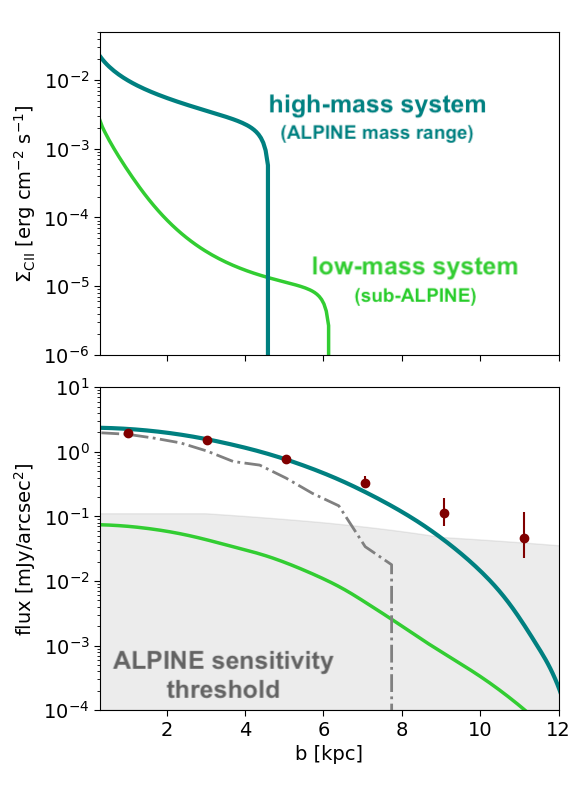}
    \caption{\textit{Top}: \CII surface density $\Sigma_\mathrm{CII}$ (eq. \ref{eq:final_intensity}) as a function of the impact parameter $b$, for the two runs considered in Sec. \ref{sec:non-det}. The low-mass system (green line) has a stellar mass of $M_*=10^9\,\msun$, corresponding to a mass loading factor $\eta=12.8$, a star formation rate $\mathrm{SFR}=2.6\,\msun\,\mathrm{yr}^{-1}$, and a circular velocity $V_c=143\,\kms$. The high mass system (teal line), instead, has $M_*=10^{10}\,\msun$, $\eta=4.7$, $\mathrm{SFR}=62\,\msun\,\mathrm{yr}^{-1}$, $V_c=218\,\kms$. \textit{Bottom}: Predicted flux for the two systems, obtained by convolving the surface brightness with the same beam as observations (dashed-dotted gray line) and converting the units to mJy/arcsec$^2$. In this way, a direct comparison with observational data is possible (the data points correspond to the system DC$396844$ and are shown only as a reference). The gray shaded region shows the ALMA sensitivity threshold for ALPINE observations.
    \label{fig:final_dep}
 	}
\end{figure}

Given that the ubiquity of galactic winds at all redshifts and their key role in shaping galaxy properties have been widely acknowledged \citep[][]{rupke2018review}, a causal connection between outflows and the formation of \CII halos seems at least plausible. However, this connection needs to be backed up by direct observational evidence of galactic winds in \CII halo-hosting systems. The best way to probe the presence of these winds is by studying the kinematics of the gas. Looking at the spectrum of the \CII line, in fact, it is possible to search for a broad component that is commonly associated with the presence of outflowing gas.
Indeed, several works have already found preliminary evidence for this. The first discovery of \CII halos in normal star-forming galaxies \citep[][]{Fujimoto19} was accompanied by a statistical indication of the presence of broad wings in the stacked \CII spectra of nine of these $z\approx 6$ galaxies \citep[][]{gallerani:2018}.
In the context of the ALPINE survey, \citet{ginolfi:2019} performed a stacking of the \CII spectra of a subsample of ALPINE galaxies, finding evidence for a broad Gaussian feature ($\mathrm{HWHM}\approx275\,\kms$) that becomes even more prominent when only highly star forming ($\mathrm{SFR}>25\,\msun\,\mathrm{yr}^{-1}$) systems are considered. 

An even more convincing sign of the direct link between halos and outflow activity comes from deep observations of the individual system HZ4 \citep{herrera2021kiloparsec}. In addition to the spatially extended \CII emission already mentioned in Sec. \ref{sec:introduction}, from the line broad wings the authors identify the presence of an outflow in two adjacent sub-regions of $\approx2\,\mathrm{kpc}$ size, extending from the central disk in the direction of the minor axis. These findings point to the presence of an active outflow, with velocities in the range $v\approx 300-400\,\kms$, launched from the inner galactic regions and reaching out the CGM, that is detected as \CII extended emission. 

The above body of evidence, although still limited in size and robustness, supports the outflow-halo model proposed in this work. Our model, in fact, is characterized by gas velocities ranging (for the best-fitting $\eta$ values) from $v\approx200\,\kms$ to $v\approx500\,\kms$ (Fig. \ref{fig:gravity}, right panel). Such values of the gas velocity are comparable to the halo virial velocity, $V_{\rm c}$ (Fig. \ref{fig:gravity}, left panel). This implies that outflows are likely unable to escape the global gravitational potential of the galaxy. Rather, the expelled gas builds up a relatively high-density and high-metallicity halo that can survive for several dynamical timescales of the parent galaxy. Also, we predict that \CII emission extends out to $r\approx15\,\mathrm{kpc}$, which is in very good agreement with observed halo sizes.

This general concordance between the wind model considered here and observational evidence for \CII halos does not rule out the possibility that alternative wind models are also compatible with the formation of halos at the high redshift. For instance, cold wind modes (for which the amount of \CIIion is expected to be significant, see \citeP20 for more details) can also be formed by ISM clouds that are entrained in the hot outflowing medium. However, despite the fact that many theoretical efforts have been directed to study the evolution of these clouds in the wind \citep[e.g., ][]{scannapieco2015launching, mccourt2015magnetized, bruggen2016launching,schneider2017hydrodynamical} --, a general consensus on the role of this mechanism as a viable cold-mode formation channel has yet to be reached. As \citet[][]{fielding2021} suggests, it is possible that both radiative cooling of hot gas and cold-cloud entrainment can play a role in determining the final multi-phase structure of the outflow. Further work is needed to investigate whether this scenario gives rise to an extended \CII emission in agreement with observations.

Two additional remarks are worth adding here. First, the presence of a halo does not necessarily imply an ongoing outflow, i.e. some halos maybe be remnants of past outflow activity and could be observed well after gas ejection from the galaxy has ended. This situation might, however, easily lead to the non-detections reported by \citeF20 and \citet[][]{spilker} as the low velocities implied make the identification of the broad wings problematic. Second, we underline that our model describes a steady-state solution (for details, see \citeP20), which implicitly requires that the outflow lasts for several \CII-halo crossing times ($\approx 50$ Myr). If the system is caught at early times, the outflow might not yet have had sufficient time to build the halo.

\subsubsection{\CII halo non-detections}\label{sec:non-det}

If SF-driven outflows are responsible for the formation of \CII halos, then it is fair to ask why halos are not observed in every starburst galaxy. In fact, one of the key findings of \citeF20 is that not all of the galaxies are \CII halo hosting systems. About $\sim 25\%$ of them ($\sim 70\%$ considering galaxies classified as "other") show no detectable signs of extended \CII emission. Interestingly, these authors find that galaxies belonging to this \quotes{no \CII halo} class tend to have lower star formation rates and stellar masses with respect to \CII halo-hosting ones. A similar conclusion was also drawn by \citet[][]{ginolfi:2019}, whose stacking revealed the presence of extended \CII emission only for the subsample of galaxies with ${\rm SFR} >25\,\msun\,\mathrm{yr}^{-1}$. 

A possible explanation for these trends is that low-mass and low-SFR galaxies cannot launch outflows that are powerful enough to transport carbon away from the galactic center. However, such hypothesis is in contrast with the fact that outflows are seen in simulations in every mass range, and are expected theoretically from the metal enrichment scenario at high-$z$ \citep[e.g.][]{dave2012,wise2014birth,pallottini2014}; even more critically, in low mass systems they might be more prominent as a result of their larger mass loading factors \citep[see Sec. \ref{sec:results} and][]{muratov2015}. 

Our model naturally offers a different interpretation for the non-detections. \textit{In low-mass systems, outflows are indeed present and transport carbon into the circumgalactic medium, but the resulting extended emission is too faint to be detected with the current level of sensitivity.} This can be showcased by considering our model with two different values of the stellar mass: $M_*=10^9\,\msun$ (labelled as \textit{low mass}) and $M_*=10^{10}\,\msun$ (\textit{high mass}). Using the $\eta-M_*$ (eq. \ref{eq:eta_mstar}) and $\eta-\mathrm{SFR}$ (eq. \ref{eq:eta_sfr}) relations, as well as the stellar mass-halo mass relation from \citet[][]{behroozi2013average}, we can express the three model parameters as a function of $M_*$ only. In particular, $M_*=10^9\,\msun$ corresponds to ($\eta$, SFR, $V_c$) = ($12.8, 2.6\,\msun\,\mathrm{yr}^{-1}, 143\,\kms$); for $M_*=10^{10}\,\msun$ we find (4.7, $62\,\msun\,\mathrm{yr}^{-1}$, $218\,\kms$).

In Fig. \ref{fig:final_dep} (upper panel), we plot the surface brightness of the \CII emission for these two runs. The high mass system has a larger surface brightness at all radii, but the profile is quite compact, because the outflow is halted by gravity at $\simeq 4.5\,\mathrm{kpc}$ from the launching site. On the other hand, the low mass one is more extended, but the absolute value of the surface brightness is 10$\times$ lower at the center, also showing a steeper radial decline. 

Concluding that larger $\eta$ values correspond to dimmer profiles may seem counterintuitive, as one might naively expect that an outflow with high $\eta$ is loaded with more material. However, the relatively shallow $\eta(M_*)$ trend in momentum-driven winds implies that it is the SFR, rather than $\eta$, that ultimately governs the mass outflow rate for a system. More massive and star-forming systems are therefore characterized by mass-loaded outflows that can be dense and bright enough to form extended halos in \CII.

This is clear by looking at the bottom panel of Fig. \ref{fig:final_dep}, where we convolve the surface brightness profiles with the ALMA beam and transform the surface brightness in a flux per unit solid angle (measured in mJy/arcsec$^2$). We have assumed the beam size, redshift, and FWHM of the system to be equal to those of DC$396844$; such choice does not affect our conclusions in any relevant way. As a reference, we also plot the data for DC$396844$ (red points), as well as the associated line sensitivity (gray shaded region) taken from \citeF20 \citep[corresponding to an integration time of $\sim 20\,\mathrm{min}$,][]{bethermin:2019}. 
The flux of the high-mass system is well above the noise level, and reaches an effective size of $r\gtrsim10\,\mathrm{kpc}$, which is partly due to beam smoothing. Thus, this system would be classified as a galaxy with a \CII halo. On the other hand, the low-mass system flux radial profile is very faint: it is slightly below the ALMA sensitivity level at the center, and it declines rapidly such that the region extending farther than the beam size has a very low surface brightness ($\sim10^{-3}\,\mathrm{mJy}/\mathrm{arcsec}^2$). 

These results are in line with observations, as a mass of $M_*=10^{10}\,\msun$ falls exactly in the range where \CII halos have been detected by the ALPINE survey (Tab. \ref{tab:alpine_sample}). The low mass system ($M_*=10^{9}\,\msun$) is instead representative of the group of low-mass/low-SFR galaxies whose \CII emission is not extended enough to be classified as a halo by \citeF20 and by the \citet[][]{ginolfi:2019} stacking. Probing the presence of \CII halos in this set of galaxies with future observations may not be trivial, as a very high sensitivity is required due to the steep decrease in surface brightness at large radii. The optimal solution could consist in improving both the sensitivity level and the angular resolution of observations, in order to constrain the emission in the peripheral region at $r\sim4-8\,\mathrm{kpc}$.

The above discussion argues in favor of a picture in which starburst-driven outflows are widespread in high-$z$ galaxies at all masses. The detection of the extended \CII emission associated with this outflow activity, however, is subject to the sensitivity level and spatial resolution of observations. We highlight the fact that this discussion is useful to interpret the general trends observed in \citeF20, but it is not able to capture the differences in the halo properties that the authors found in systems with very similar stellar masses and star formation rates. We expect that other factors, such as time-dependent effects discussed above and not included in our simplified model, may contribute to create these differences.

\subsubsection{The role of the escape fraction}

Finally, we also point out that the arguments presented here are somewhat sensitive to the assumptions concerning the ionizing photon escape fraction, $\fesc\ll1$ (see Sec. \ref{sec:model}). A negligible escape fraction is a key requirement for our model to be compatible with the observed halo luminosities, as photons emitted from the central galaxies drastically reduce the amount of neutral gas in the outflow. However, the conclusion $\fesc\sim 0$ holds true only for galaxies that have been observed to host a \CII.
Indeed, since $\fesc$ has a major impact on the formation of \CII halos, its variation in different mass ranges \citep[e.g.,][]{xu2016galaxy,wise2014birth} could also be responsible for the absence of observed \CII halos in low-mass systems. If this is the case, then we foresee the opportunity of using the presence of \CII halos as a tracer of $\fesc$ at high redshifts. The James Webb Space Telescope (JWST) could help us testing this by determining the value of the escape fraction in the low-mass, low-luminosity range \citep[][]{chrisholm}.

A small but non-negligible value of $\fesc$ is expected in low masses galaxy, which are the main driver of the first stages of Reionization \citep[e.g.,][]{ma2020no}. However, the presence of neutral gas in the CGM is not in contrast with the leakage of ionizing photons from the central galaxy, as photons preferentially escape through low-column density paths that cover only a small fraction of the solid angle \citep[e.g.][]{stern}. For this reason, we can speculate that high $\fesc$ channels provide a sufficient amount of ionizing photons to reionize the IGM, while, at the same time, the thermal/ionization structure of the outflows is preserved in most of the volume. This possibility can be explored within our model by dropping the assumption of spherical symmetry. We defer this study to future work.

\subsection{Outflow mass loading factors} \label{sec:loading_factors}

It is interesting to compare the mass loading factors we obtain in Sec. \ref{sec:results} ($4.3 < \eta < 6.8$) with observational estimates, both at low and high redshifts. However, it is important to highlight the fact that our analysis focuses only on the systems showing clear evidences of a \CII halo. Therefore, the values of $\eta$ we find are representative only of this special class of high-$z$ galaxies, which may be a biased sub-sample of the whole high-$z$, star-forming galaxies population. 

Moreover, although the outflow properties largely depend on $\eta$ as shown by \citet[][]{Thompson16}, other model parameters such as the wind energy coupling and the number of supernovae per unit stellar mass ($\alpha$ and $\nu$ in \citeP20, respectively) may also play a role in determining the final extension and property of the resulting \CII halo. In \citeP20, we assumed $\alpha=1$ \citep[consistently with observational estimates from][]{strickland2009supernova} and $\nu = 0.01\, \msun^{-1}$. Although reasonable, different choices for these parameters may affect the optimal $\eta$ range found in this work. 

With these caveats in mind, we briefly review other estimates in the literature. A relevant work by \citet{Heckman15} uses UV absorption lines to study the properties of starburst-driven outflows in local starburst galaxies. They find that $\eta$ weakly anti-correlates with both $\mathrm{SFR}$ and $V_{\rm c}$ (or, equivalently, $M_*$), and $1 \lesssim \eta \lesssim 4$. This range is lower than the one we find in our study. This discrepancy may be explained by an evolution of $\eta$ with redshift \citep[e.g.,][]{nelson_2019}, which is expected given the different properties of the ISM in high-$z$ galaxies \citep[][]{Pallottini22}. Other estimates, however, find significantly larger $\eta$ values, which are compatible with our conclusions. For instance, using H$\alpha$ emission lines originating in the CGM of local galaxies, \citet{zhang2021empirical} measure $\eta$ in different ranges of halo masses. In the range relevant for the ALPINE galaxies studied here, they find $2 \lesssim \eta \lesssim 7$, in very good agreement with the values found here. 

The \citet{ginolfi:2019} analysis of stacked ALPINE systems gives relatively low values of $\eta$ ($\approx 0.5-2$). However, \citet{herrera2021kiloparsec} -- using deep observations of the galaxy HZ4 -- find similar $\eta$ values only when considering the \textit{total} SFR of the galaxy. According to their study, only a sub-region of the galaxy is driving the wind; when computing the mass loading factor by using the local star formation rate of the regions showing signs of outflow activity, they find $\eta = 3-6$ (grey shaded box in Fig. \ref{fig:final}). This estimate is compatible with our results, and suggests that measurements of the mass loading factors could be underestimated when considering the entire galaxy as the outflow launching site. Deeper observations are needed to determine whether this is the case for the majority of galaxies in the ALPINE sample.

\section{Summary} \label{sec_conclusions}

In this work, we showed that the recently discovered, extended ($\approx 10-15\,\mathrm{kpc}$) \CII emitting halos around high-$z$ galaxies can be produced by ongoing (or past) outflow activity. We improved and extended the model presented in \citeP20, by adopting a more realistic (DM+baryons) density distribution, and accounting for the effects of cosmic microwave background (CMB) suppression. Our model generates synthetic \CII radial profiles depending on three key parameters: the outflow mass loading factor ($\eta$), the galaxy star formation rate ($\mathrm{SFR}$), and the parent DM halo circular velocity at virial radius ($V_{\rm c}$). Using a MCMC-based approach, we compare our model with individual targets in the ALPINE survey \citep[][]{Fujimoto:2020qzo}, and infer the values of these parameters for each \CII halo-hosting system. The results are summarized as follows. 

\begin{itemize}
    \item Our improved model is in broad agreement with the results from \citeP20: outflows launched by SN explosions in the centre of galaxies expand in CGM with velocities in the range $200-500\,\kms$. Within the first kpc, the gas catastrophically cools to $T \approx {\rm few} \times 100$ K, and then it is slowly heated up again to $T\approx10^3\,\mathrm{K}$ by the cosmic UV background. Such conditions are favorable for the formation of \CIIion ions. If the escape fraction of ionizing photons from the parent galaxy is low ($\simlt 1$\%), the outflows remain largely neutral, with a significant \CIIion abundance at any stage of their evolution.
    \item The halo gravitational pull slows down the outflow, typically stopping it completely at $r\approx 10-15\,\mathrm{kpc}$. These values are perfectly compatible with the extension of the observed \CII emission. Gravitational effects are particularly important for large $\eta$ values.
    \item CMB can suppress the \CII emission of the halo by a factor $2-4$. Even accounting for this suppression the observed \CII luminosity ($L_\mathrm{CII}\approx 5-15\times10^8\,\mathrm{L}_\odot$) can be recovered.
    \item By comparing our model predictions with observational data of individual galaxies in the ALMA ALPINE survey, we show that (a) detected \CII halos are a natural by-product of starburst-driven outflows; (b) the mass loading factors of these outflows have values in the range $4 \simlt \eta \simlt 7$ and scalings compatible with the momentum-driven hypothesis ($\eta \propto M_*^{-0.43}$).
    \item Our model (and simulations in the literature) suggest that outflows are widespread phenomena in high-$z$ galaxies. However, in low-mass systems the halo extended \CII emission is likely too faint to be detected with the current sensitivity levels.
\end{itemize}

The fact that the extended \CII halos surface brightness can be successfully fit by our model does not entail that outflows are the only possible explanation. Other possibilities need to be considered. These include the presence of faint satellite galaxies and disturbed morphologies due to ongoing merger activity. 
While such alternative scenarios seem plausible, given that mergers and satellites are common features in simulations at high redshift \citep[][]{kohandel:2019, Gelli2020}, note that \citet[][]{Fujimoto19} discard galaxies with disturbed morphologies from the sample used in their analysis. Also, the increasing trend of the \CII luminosity-SFR surface density ratio with radius would imply a non-physical \CII emissivity in satellites.
It is difficult to confirm the dynamical effect of merger/satellites from current observations \citep{rizzo:2022}. Thus, dedicated ALMA observations, such as the ongoing CRISTAL Large Program (PI: Herrera-Camus), can probe the role of outflows and alternative halo formation scenarios either by spatially resolving the presence of satellites and disturbed morphologies, or by finding direct signs of outflows in the \CII line spectra of halo-hosting systems (Sec. \ref{sec:discussion}). The James Webb Space Telescope can also be used to resolve the halo regions in $z\sim4-6$ galaxies and to, e.g., gauge the contribution of satellite galaxies to the total \CII extended emission \citep[][]{Gelli2021}.

In any case, exploring theoretically multiple scenarios is especially appropriate considering that, despite its success, the model presented here contains several limitations and hypotheses that will need to be further refined in the future.
For instance, the assumption of spherical symmetry prevents a thorough study of more realistic halos morphologies revealed by observations. In the same way, the steady-state approach hinders a careful assessment of the halo evolution under the influence of the dark matter gravitational potential. Gas turbulence and interactions with the external CGM/IGM environment may also affect the wind evolution and change the density structure of the gas, which is directly connected to the final \CII emission. Furthermore, non-equilibrium cooling and recombination should also be included, as they can significantly alter the final temperature profiles \citep[][]{gray2019catastrophic,danehkar_2021}.

Despite the intrinsic limitations of our model, it is very encouraging to find such a good level of agreement with data within a clear, physically-motivated framework. 
On this basis, we concluded that observed halos are the likely smoking gun of early CGM enrichment caused by supernova-driven outflows in primordial galaxies.

\section*{Acknowledgements}
We are grateful to Seiji Fujimoto for sharing his study on ALPINE data with us. 
EP is grateful to Matus Rybak and Kirsty Butler for fruitful discussion. 
AF, AP, LS, MK, SC acknowledge support from the ERC Advanced Grant INTERSTELLAR H2020/740120 (PI: Ferrara). Any dissemination of results must indicate that it reflects only the author’s view and that the Commission is not responsible for any use that may be made of the information it contains.
Partial support from the Carl Friedrich von Siemens-Forschungspreis der Alexander von Humboldt-Stiftung Research Award is kindly acknowledged (AF).
We gratefully acknowledge computational resources of the Center for High Performance Computing (CHPC) at SNS.
We acknowledge use of the Python programming language \citep{VanRossum1991}, Astropy \citep{astropy}, \code{emcee} \citep{foreman2013emcee}, Corner \citep[][]{corner}, Matplotlib \citep{Hunter2007}, NumPy \citep{VanDerWalt2011}, and SciPy \citep{scipyref}.


\subsection*{Data availability}

The derived data generated in this research will be shared on reasonable requests to the corresponding author.

\bibliographystyle{mnras}
\bibliography{biblio,codes} 


\appendix

\section{Details on the CMB suppression} \label{sec:cmb_appendix}

We study the effects of CMB photons on the \CIIion level populations. Specifically, we focus on the levels that give rise to the \CII $158\,\mu\mathrm{m}$ ($^2P_{3/2} \rightarrow ^2P_{1/2}$). We consider these two levels as a closed system, and, assuming statistical equilibrium, we write the balance equation for these two levels as \citep[][]{gong2012, dacunha2013}:
\begin{equation}
    n_l (n_e\,\gamma_{lu} + \,B_{lu}\, I(\nu)) + P_{lu}^\mathrm{(UV)} = n_u(n_e\,\gamma_{ul} + A_{ul} + B_{ul} \,I(\nu)) + P_{ul}^\mathrm{(UV)} \label{eq:level_balance}
\end{equation}
In this relation, $n_l$ and $n_u$ are the numeric densities of the two levels, and $n_e$ is the density of free electrons, i.e. the main collisional partners for \CII in the range of densities we are interested in \citep[][]{vallini2015}. $A_{ul}$, $B_{ul,lu}$, the Einstein coefficients, $\gamma_{ul,lu}$ the collisional excitation rates, and $I(\nu) = B(\nu, z)$ is the (black-body) specific intensity of the CMB radiation. 

Finally, $P_{ul,lu}^\mathrm{(UV)}$ are UV excitation and de-excitation rates. These account for the \textit{UV pumping} effect: in the presence of FUV radiation at $1330$ \AA, electrons can be pumped from the $^2P_{3/2}$ ($^2P_{1/2}$) level to $^2D_{3/2}$ at $1335.66$ \AA \,($1334.53$ \AA). This can lead to the \CII fine structure transitions $^2D_{3/2}\rightarrow \,^2 P_{3/2} \rightarrow \,^2 P_{1/2}$, resulting in a mixing of the levels of the \CIIion doublet. The UV rates parametrize this effect in the context of a two-level treatment, and are given by the following expressions \citep{field}:
\begin{align}
      P_{ul}^\mathrm{(UV)} &= \frac{g_k}{g_u}\frac{A_{kl}A_{ku}}{A_{kl}+A_{ku}}\left(\frac{c^2 I_\mathrm{UV}(\nu_{ku})}{2h\nu_{ku}^3}\right)\\
    P_{lu}^\mathrm{(UV)} &= \frac{g_k}{g_l}\frac{A_{kl}A_{ku}}{A_{kl}+A_{ku}}\left(\frac{c^2 I_\mathrm{UV}(\nu_{kl})}{2h\nu_{kl}^3}\right)
   \end{align}
where $I_\mathrm{UV}$ is the UV radiation intensity, and $k$ is the level $^2D_{3/2}$, which has degeneracy $g_k = 4$; $A_{kl} = 2.41\times10^8\,\mathrm{s}^{-1}$ ($\nu_{kl}$) and $A_{ku} = 4.76\times 10^7\,\mathrm{s}^{-1}$ ($\nu_{ku}$) are the Einstein
coefficients (frequencies) of the $^2D_{3/2}\rightarrow \,^2P_{1/2}$ and $^2D_{3/2} \rightarrow \,^2P_{3/2}$ transitions, respectively.

The ratio between the UV rates determines the \textit{UV color temperature}, defined as (for the case of interest, $T_\mathrm{UV}\approx10^4\,\mathrm{K}$): 
 \begin{equation}
     T_\mathrm{UV} = T_*\log\left(\frac{g_u P_{ul}^\mathrm{(UV)}}{g_l P_{lu}^\mathrm{(UV)}}\right) \,, \label{eq:uv_color}
 \end{equation}
where $T_*\approx 91\,\mathrm{K}$ is the equivalent temperature of the \CIIion transition.

It is also convenient to introduce the \textit{excitation temperature} of the transition, which is defined in a similar way:
\begin{equation}
     T_\mathrm{exc} = T_*\log\left(\frac{g_u n_l}{g_l n_u}\right)  \label{eq:exc_temp}
 \end{equation}
Using the relations linking the three Einstein coefficients, and the two collisional excitation rates, we can rewrite eq. \ref{eq:level_balance} and obtain the following expression for $T_\mathrm{exc}$:
\begin{equation}
    \frac{T_*}{T_\mathrm{exc}} = \log\left( \frac{A_{ul}\left(1+c^2I(\nu_*)/2h\nu_*^3\right))+n_e\gamma_{ul}+P_{ul}^\mathrm{(UV)}}{A_{ul}\,c^2I(\nu_*)/2h\nu_*^3+n_e\gamma_{ul}\,e^{-T_*/T}+P_{lu}^\mathrm{(UV)}} \right)\,\label{eq:exc_temp}
\end{equation}

Once $T_\mathrm{exc}$ is known, it is possible to compute the CMB suppression factor according to eq. \ref{eq:suppression} (Sec. \ref{sec:CMB_suppression}).

\section{Details on the model-data comparison}\label{sec:data_analysis_details}

To compare our model with data, we adopt a Bayesian framework, i.e. we compute the likelihood and posterior distributions. The former quantifies the model-data agreement; the latter determines the best-fit values of the model parameters. 

We denote the parameter set with $\Theta$, and our model with $m(b; \Theta)$, where $b$ is the impact parameter of the surface brightness profiles. We consider as parameters (a) the mass loading factor $\eta$, (b) the star formation rate $\mathrm{SFR}$, and (c) the circular velocity $V_{\rm c}$. In principle, the redshift, $z$, is another parameter. However, for our sample high-quality spectroscopic redshifts are available (Tab. \ref{tab:alpine_sample}). Hence, we set $z=z_\mathrm{CII}$, and limit our parameter space to $\Theta = (\eta, \mathrm{SFR}, V_{\rm c})$.

The likelihood of a set of $n$ data points $d=\{y_i(b_i)\pm \sigma_i\}$ is
\begin{equation}
    \mathcal{L}(d\,|\Theta, m) = \exp\left(-\sum_{i=1}^n \frac{(y_i - m(b_i;\Theta))^2}{\sigma_i^2}\right). \label{eq:likelihood}
\end{equation}

The posterior can be expressed in terms of the likelihood and the prior distributions. Our \textit{a-priori} knowledge of the parameters varies. The mass loading factor is largely unconstrained both theoretically and observationally; even its order of magnitude is quite uncertain, ranging from $\eta \lesssim 0.1$ to $\eta \gtrsim 10$ \citep[e.g.,][]{muratov2015}. For this reason, the most suitable prior for $\eta$ is a logarithmic prior (i.e., a uniform prior for $\log \eta$). Therefore, from now on we work with the parameter $\log \eta$, assigning to it a uniform distribution in $-1 < \log\eta < 1.5$. 

On the other hand, both the $\mathrm{SFR}$ and $V_{\rm c}$ are somewhat constrained by the ancillary data reported in Tab. \ref{tab:alpine_sample}.
In principle, we could set these parameters to the median values inferred from observations. However, the inferred star formation rates and virial masses are relatively uncertain (errors $\simlt 50\%$). Given that our model turns out to be sensitive to relatively small changes in these parameters, we include them in the analysis, using the information we have from observations to set the priors properly. Given that the uncertainties of $\mathrm{SFR}$ and $V_{\rm c}$ are roughly symmetric in logarithmic space, we choose log-normal priors for both of these quantities. The mean of the distribution is taken to be equal to the (logarithm of) their measured values, and the standard deviation is set to half of the total (upper+lower) relative uncertainty on these values.

Overall, the prior distribution we consider is null outside the region $-1 < \log\eta < 1.5$; inside this region it is proportional to 
\begin{equation}
\begin{split}
    \pi(\Theta|m) \propto\exp\bigg(&-\frac{\left(\log_{10}\mathrm{SFR}-\log_{10}\mathrm{SFR}_i\right)^2}{2\sigma_{\log,\mathrm{SFR},i}^2}-\\
    &-\frac{\left(\log_{10}V_{\rm c}-\log_{10}V_{{\rm c},i}\right)^2}{2\sigma_{\log,V_{\rm c},i}^2}\bigg),\\
\end{split}
\end{equation}
where $\log_{10} \mathrm{SFR}_i \pm \sigma_{\log,\mathrm{SFR},i}$, and $\log_{10} V_{{\rm c},i} \pm \sigma_{\log,V_{\rm c},i}$ refer to the values displayed in Tab. \ref{tab:alpine_sample}.

Once the priors are set, the posterior distribution is well-defined, and it can be explored with a sampling algorithm such as a Markov Chain Monte Carlo \citep[MCMC,][]{metropolis, hastings}. Here, we use the well-known code \code{emcee} \citep{foreman2013emcee} to run an MCMC using the affine-invariant ensemble prescription to generate new samples \citep{goodman_prescription}. For each system considered (Tab. \ref{tab:alpine_sample}), we run the MCMC algorithm by placing $m=48$ walkers distributed randomly in the parameter space and evolving them for $N>10^5$ steps. We set the final number of steps so that our chain is at least 100 times longer than the autocorrelation time $\tau$ \citep[see e.g.,][]{sharma2017markov}. This is easily achieved for the unimodal posteriors in Figure \ref{fig:corner_2}, while it takes longer integration times for some of the systems showing a bi-modal behavior (i.e., DC$630594$, DC$689613$, DC$881725$). As the final chain is produced, we discard the first $k>10^3$ elements to account for the burn-in phase, and we thin the chain considering only one element every $\tau$ steps in order to account for autocorrelations.

\section{Results for the whole sample} \label{sec:other_sample}

\begin{figure*}
	\centering
	\includegraphics[width=0.42\textwidth]{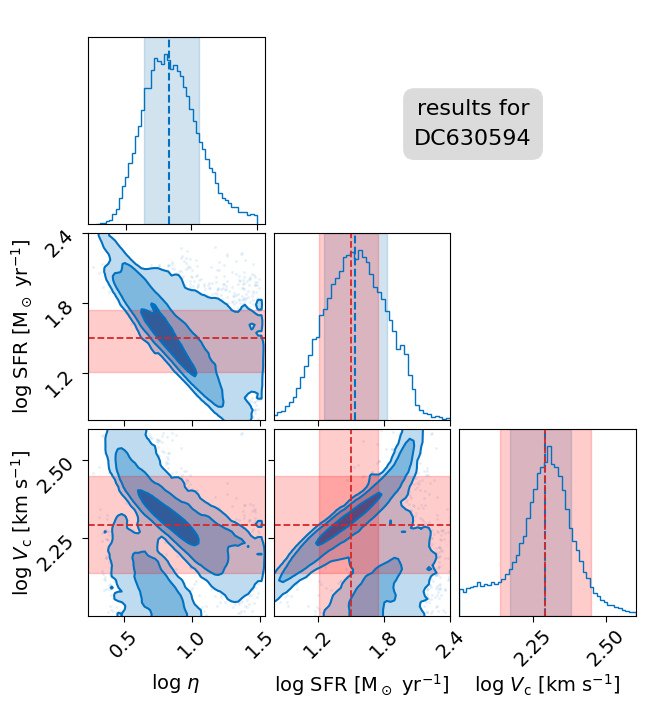}
	\includegraphics[width=0.42\textwidth]{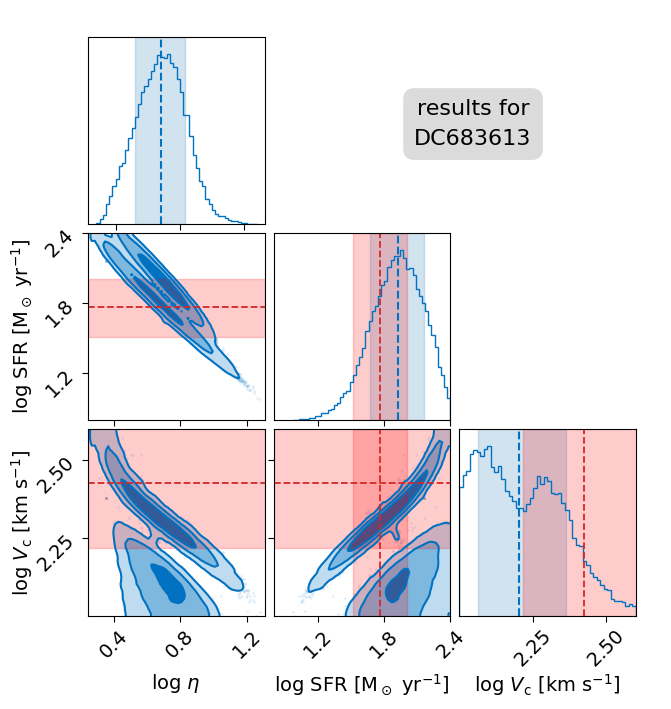}
	\includegraphics[width=0.42\textwidth]{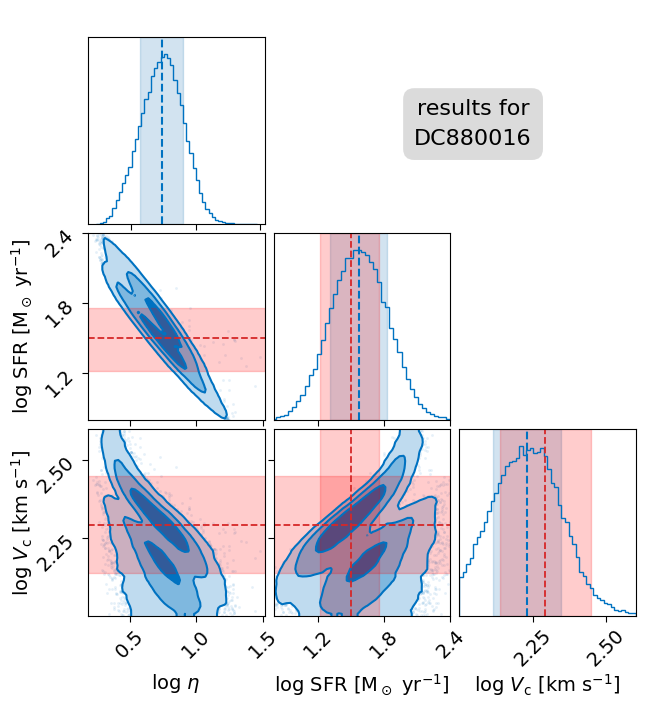}
	\includegraphics[width=0.42\textwidth]{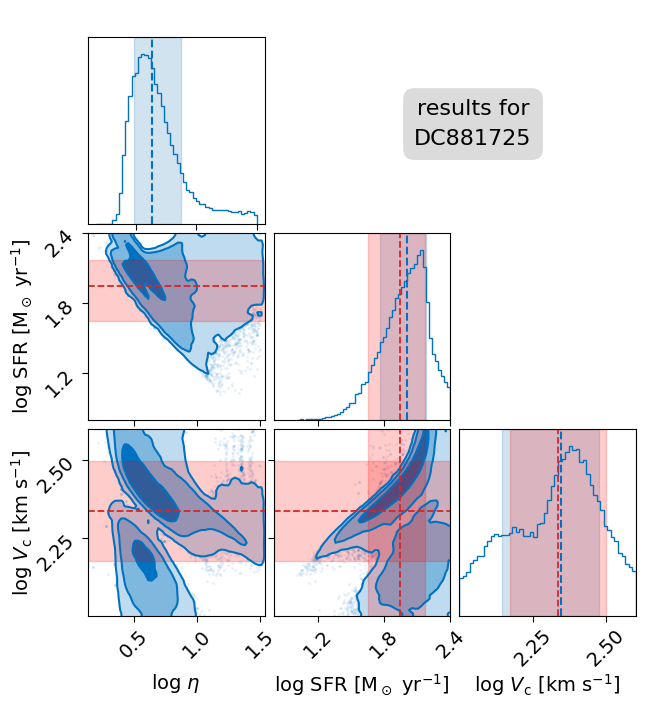}
	\includegraphics[width=0.42\textwidth]{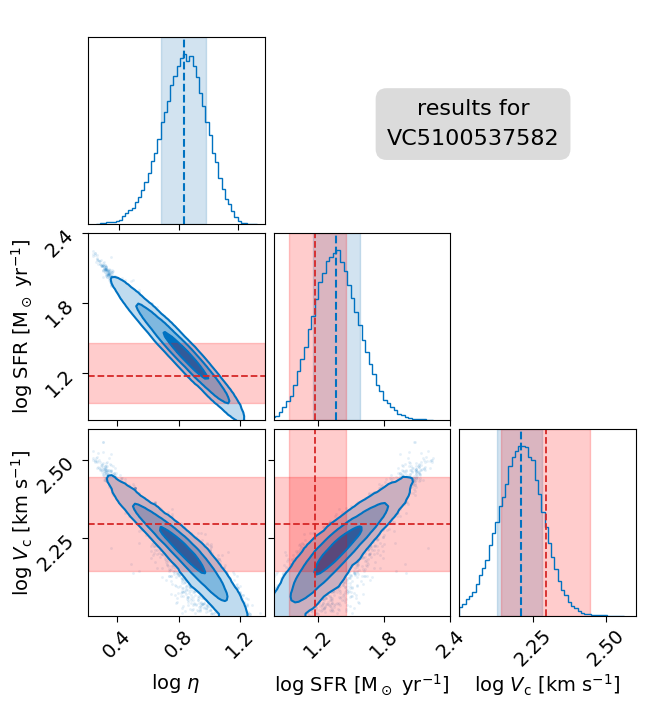}
	\includegraphics[width=0.42\textwidth]{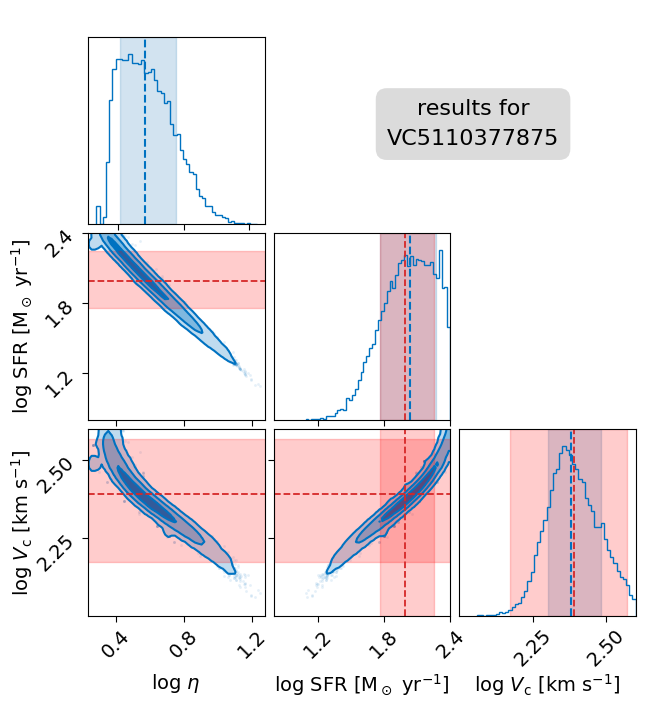}
	\caption{Corner plots for all the systems but DC$396844$ (shown in the main text). Details can be found in Fig. \ref{fig:corner_1} (left panel).
	\label{fig:corner_2}
	}
\end{figure*}

In Fig. \ref{fig:corner_2}, we show the posterior distributions for all the systems considered in Tab. \ref{tab:alpine_sample} with the exception of DC$396844$, which is discussed at length in the main text (see Fig. \ref{fig:corner_1}).

\bsp	
\label{lastpage}

\end{document}